\documentclass{llncs}
\usepackage{llncsdoc}
\usepackage{stmaryrd}
\usepackage[color,all,poly,matrix,arrow]{xy}
\usepackage{graphicx}
\usepackage{amssymb, amsmath,amscd}

\newcommand{\LTO}[1]{\overset{\sf{#1}}{\bullet}}
\newcommand{\DTO}[1]{\underset{\sf{#1}}{\bullet}}

\def\Inst{{\bf Inst}}
\def\iinst{{{\text \textendash}\bf \Inst}}
\def\taking{\colon}
\def\tn{\textnormal}

\begin{document}

\title{Functorial Data Migration: \\ From Theory to Practice}

\author{Ryan Wisnesky\inst{1} \and David I. Spivak\inst{1} \and Patrick Schultz\inst{1} \and \\ Eswaran Subrahmanian\inst{2} }
\institute{Massachusetts Institute of Technology\footnote{Work supported by ONR grant N000141310260, AFOSR grant FA9550-14-1-0031, and NASA grant NNH13ZEA001N-SSAT.} \and Carnegie Mellon University and the National Institute of Standards and Technology (NIST)}

\maketitle

\begin{abstract}
In this paper we describe a functorial data migration~\cite{Spivak:2012:FDM:2324905.2325108} scenario about the manufacturing service capability of a distributed supply chain.   The scenario is a category-theoretic analog of an ontology-based ``semantic enrichment'' scenario~\cite{serm} developed at the National Institute of Standards and Technology (NIST).  The scenario is presented using, and is included with, the open-source FQL tool, available for download at {\sf categoricaldata.net/fql.html}. 
\end{abstract}
\section{Introduction to Functorial Data Migration}
In the {\it functorial data model}, which originated with Rosebrugh and others in the late 1990s~\cite{Fleming02adatabase}, a database schema is a finitely presented category~\cite{BW} (essentially, a directed multi-graph and path equality constraints) and a database instance on a schema $S$ is a set-valued functor from $S$ (essentially, a set of tables).  The database instances on a schema $S$ constitute a category, denoted $S\iinst$, and a functor $F\taking S \to T$ between schemas $S$ and $T$ induces three adjoint data migration functors: $\Delta_F\taking T\iinst \to S\iinst$, defined as $\Delta_F(I) := I \circ F$, and the left and right adjoints to $\Delta_F$, respectively: $\Sigma_F\taking S\iinst \to T\iinst$ and $\Pi_F\taking S\iinst \to T\iinst$.  These data migration functors provide a category-theoretic alternative to traditional, set-theoretic operations for information integration such as SQL and the chase~\cite{FKMP05}.   

 We have developed a simple algebraic query language for the functorial data model, {\it FQL} (for Functorial Query Language), as well as a corresponding integrated development environment (IDE), the {\it FQL IDE}.  The FQL IDE is an FQL code editor, a FQL $\leftrightarrow$ SQL translator, a FQL execution engine, and a data visualization tool designed in the spirit of the schema-mapping tool Clio~\cite{haas:clio}.  The FQL IDE is open source, written in java, and available for download at  {\sf categoricaldata.net/fql.html}. In this paper, we demonstrate how the FQL IDE is used in practice by describing an example data migration scenario developed in collaboration with the National Institute of Standards and Technology (NIST).
\newpage
{\bf Remark.} Rosebrugh et al's original model~\cite{Fleming02adatabase} has a number of theoretical issues that prevent it from being used directly as a basis for information integration.  First, Rosebrugh's model cannot store meaningful data such as strings and integers; it can only store meaningless identifiers (IDs).  Second, Rosebrugh's model cannot interoperate with SQL.  Hence, FQL is actually based on an extension of Rosebrugh's model, described in~\cite{relfound}.  The exact definition of this extension does not matter for the purposes of this paper.

\section{An Enrichment Scenario}

The example described in this paper is an FQL analog of a ``semantic enrichment'' scenario developed at NIST and published as~\cite{serm}.  In this scenario, a database (called {\it Portal A} in~\cite{serm}) contains information about equipment, including the capabilities of such equipment; for example, that a particular machine $m$ can drill holes as small as .5cm in metal.  The goal of the scenario is to ``enrich'' Portal A's data with additional 3rd party information about materials, so that, for example, Portal A's data also contains the fact that $m$ can drill holes in iron, because iron is a kind of metal.

  In~\cite{serm}, Portal A's database is a Microsoft Access database, the 3rd party enriching information about materials is an OWL (Web Ontology Language) ontology, and enrichment is done by invoking a black-box OWL reasoner on an input query, Portal A's data, the OWL ontology about materials, and an OWL ontology relating portal A's vocabulary (e.g., ``iron'') and the material ontology vocabulary (e.g., ``ferrous''). 
In this paper, we simplify this scenario as follows: we assume Portal A's data is given as a SQL database, that the ontology about materials is simply an ``is-a" parenthood function, and that the correspondence between Portal A's vocabulary and the is-a hierarchy vocabulary is a ``synonyms'' relation between sets of words.

  Our FQL development consists of three main steps:
  \begin{enumerate}
\item First, we import Portal A's data, the is-a hierarchy, and the synonyms into FQL.  (Section 2.1)
\item Second, we transitively close the is-a hierarchy, join it with the synonymns relation, and then join the result to Portal A's data. (Section 2.2)
\item Finally, we test the result of our enrichment on a particular query (query 1 from~\cite{serm}).  This query gives additional results on the enriched data, which demonstrates that FQL can be used to do semantic enrichment along the lines described in~\cite{serm}. (Section 2.3)
\end{enumerate}
 Although we only have space to sketch the outline of the development, the entire development -- about 2000 lines of FQL code, 1800 lines of which are schema and data definitions -- is included as a built-in example in the FQL IDE.\footnote{There are three variants of the FQL IDE, each of which implements a slightly different language.  The example in this paper is ``P NIST Full'' in the ``FPQL IDE''.}  
\subsection{Step 1: Import relational data}

\begin{figure}
\begin{center}
\includegraphics[width=4.7in]{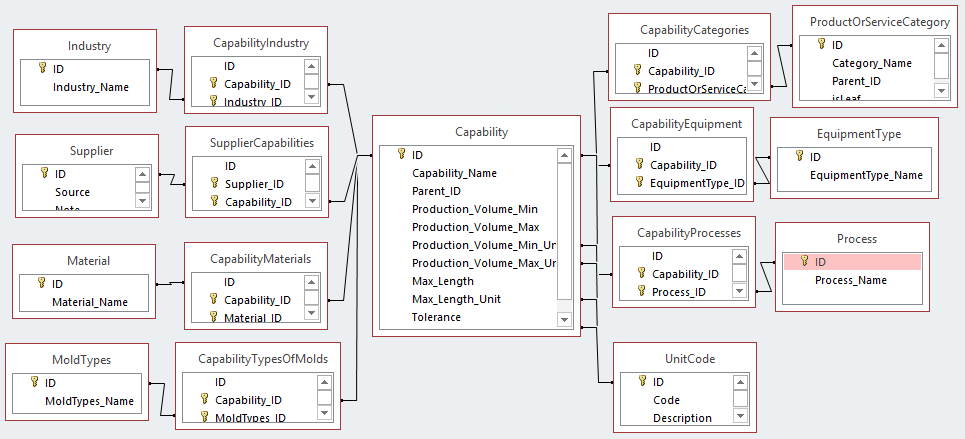}
\end{center}
\vspace{-.2in}
\caption{Schema for Portal A}
\label{schema}
\end{figure}

The schema (Figure~\ref{schema}) for Portal A's data (Figure~\ref{sql}) is a SQL schema in categorical normal form~\cite{Spivak:2012:FDM:2324905.2325108}: every table consists of a distinguished (primary key) ID column, a set of ``attribute'' columns whose values contain strings or integers, and a set of foreign key columns whose values contain IDs that refer to other tables.  Consequently, Portal A's schema can be regarded as the presentation of a category: the objects of the category are the table names and type names, and the arrows between objects are the foreign key or attribute columns in the schema.  An instance on Portal A's schema, which physically is a set of relations, can then be regarded as a set-valued functor.   

\begin{figure}
\begin{center}
\includegraphics[width=5in]{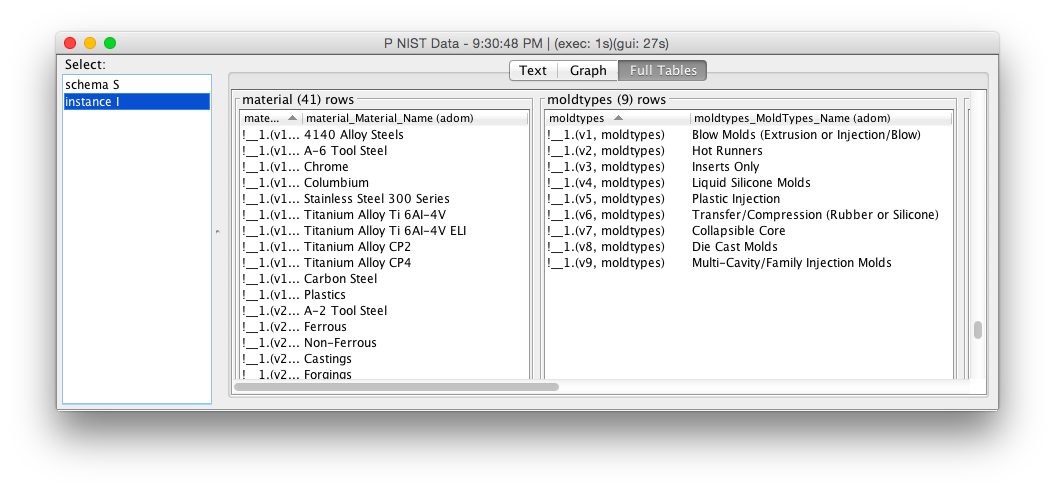}
\end{center}
\vspace{-.4in}
\caption{Portal A data displayed in the FQL IDE}
\label{portala}
\end{figure}

The actual Portal A schema as visualized in Microsoft Access is shown in Figure~\ref{schema}, and a snippet of the SQL commands defining the Portal A data are shown in Figure~\ref{sql}.  The FQL IDE imports these SQL commands and emits corresponding FQL code that defines an FQL schema and an FQL instance on that schema.  A portion of Portal A's data, as displayed in the FQL IDE, is shown in Figure~\ref{portala}.

\vspace*{-.4in}
\begin{figure}
\begin{footnotesize}
\begin{verbatim}
CREATE TABLE unitcode (
  id INT PRIMARY KEY, Code VARCHAR(255), Description VARCHAR(255)
);
INSERT INTO unitcode VALUES 
(1,"EA","Each part/piece count"),
(2,"Thousands","1000 parts/pieces count"),
(3,"Inch","Length measure in inches"),
(4,"mm","Length measure in millimeters"),
(5,"cm","Length measure in centimeters"); 
\end{verbatim}
\end{footnotesize}
\vspace{-.1in}
\caption{Snippet of SQL for Portal A}
\label{sql}
\end{figure}

\vspace{.2in}
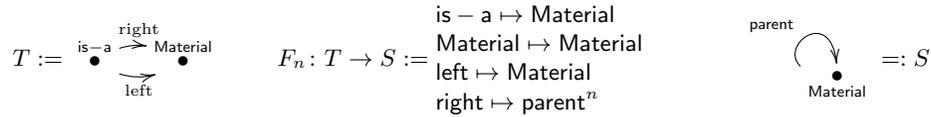
\begin{figure}[h]
\begin{align*}
 T := \parbox{.8in}{\xymatrix@=10pt{\LTO{is-a}  \ar@/^/[r]^{\tn{right}} \ar@/_/[r]_{\tn{left}}  &  \LTO{Material}}}
\hspace{.3in}
F_n \taking T \to S := \parbox{1.4in}{${\sf is-a} \mapsto {\sf Material}$ \\ ${\sf Material} \mapsto {\sf Material}$ \\ ${\sf left} \mapsto {\sf Material}$ \\ ${\sf right} \mapsto {\sf parent}^n$}
\hspace{.2in}
\parbox{.8in}{\xymatrix@=10pt{\DTO{Material}\ar@(l,u)[]^{\sf{parent}}}} =: S
\end{align*} 
\vspace{-.1in}
\caption{The reflexive transitive closure of a function $I$ is $\Delta_{F_0}(I) \cup \Delta_{F_1}(I) \cup \ldots$} 
\label{span}
\end{figure}
\vspace{.1in}

Two additional inputs are specified in the original scenario~\cite{serm}: an OWL ontology $X$ containing myriad facts about materials (e.g., steel is a metal), and an OWL ontology relating the vocabulary used by $X$ (e.g., ``ferrous'') to the vocabulary used by portal A (e.g., ``iron'').  At present we do not have a good understanding about how OWL relates to FQL.  So, we went through these ontologies by hand and stripped out relevant data.  The result was
\begin{itemize}
\item a (total) function ${\sf parent} : O \to O$, where set $O$ is the set of words from the ontology,  and 
\item a synonyms relation ${\sf syn} \subset O \times N$ where $N$ is the set of words from Portal A.  We do not require that  {\sf syn} be an equivalence relation, and for our particular data, it is not.
\end{itemize}
We encode the {\sf parent} function as an instance on the $S$ schema in Figure~\ref{span}, and because it turns out that $N$ and $O$ are disjoint, we can encode the {\sf syn} relation as an instance on the $T$ in Figure~\ref{span} by treating the target node of both edges as representing $N \cup O$.  If $N$ and $O$ were not disjoint, we would need to use a span~\cite{BW} schema with three, rather than two, nodes to encode the {\sf syn} relation, but our development would be mostly the same.  
\newpage
\subsection{Step 2: Process imported data}

\begin{figure}
\begin{center}
\includegraphics[width=4.2in]{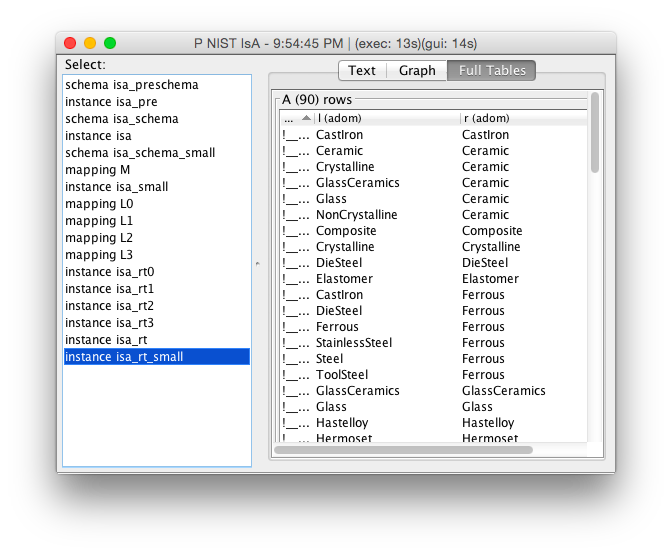}
\end{center}
\vspace{-.5in}
\caption{Transitively closed ``is-a'' relation displayed in the FQL IDE}
\label{relation}
\end{figure}

We enrich Portal A's data using the $\Sigma,\Delta,\Pi$ data migrations (defined in Section 1) as follows.  We begin by computing the reflexive, transitive closure of the parenthood function, resulting in an {\sf isa} relation.  To do this, we define, in Figure~\ref{span}, for each natural number $n$, a functor $F_n \taking T \to S$ from the schema for a relation ($T$) to the schema for a function ($S$).  Given an $S-$instance (e.g., the {\sf parent} function) $I$, $\Delta_{F_n}(I)$ computes, as a $T-$instance (i.e., relation), the $n$-ary composition of $I$, i.e., $I^n$, with the $0$-th composition being the reflexive closure of $I$.  The reflexive transitive closure of $I$ is then the union $\Delta_{F_0}(I) \cup \Delta_{F_1}(I) \cup \ldots$.  For this example, we used $n=3$.  Taking the union of two instances on the same schema is a built-in FQL primitive.\footnote{Technically, FQL has two primitives, disjoint union and relationalization (which equates IDs that are ``observationally equivalent'').  For SQL data, such as in this example, disjoint union followed by relationalization implements union.}  A portion of the resulting {\sf isa} relation, as displayed in the FQL IDE, is shown in Figure~\ref{relation}.

\vspace{-.2in}
\begin{figure}
\begin{center}
\includegraphics[width=4.2in]{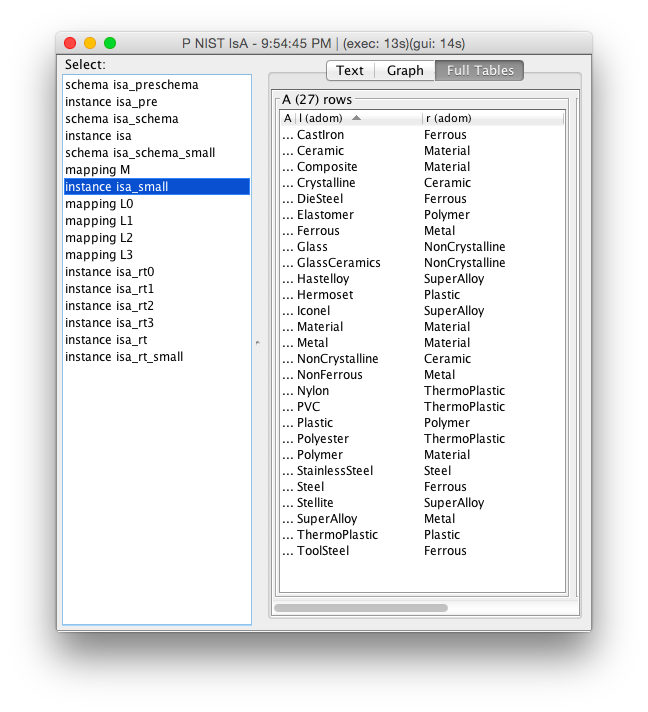}
\end{center}
\vspace{-.5in}
\caption{Initial ``is-a'' parent function displayed in the FQL IDE}
\label{function}
\end{figure}

\newpage
We now have a relation ($T-$instance) ${\sf isa} \subset O \times O$, where $O$ is the set of words from the materials ontology, and we have a relation ($T-$instance) ${\sf syn} \subset O \times N$, where $N$ is the set of words from Portal A.  We next compute a translation of {\sf isa} to use words from Portal A by joining {\sf isa} with {\sf syn} resulting in a new relation ($T$-instance) ${\sf isa''} \subset N \times N$; formally, we are computing $op({\sf syn}) ; {\sf isa} ; {\sf syn}$, where ``;`` denotes relation composition and $(x,y) \in op(R)$ if and only if $(y,x) \in R$.  Finally, we must compute the reflexive transitive closure of {\sf isa''}, which we will denote {\sf isa'}.

 To specify how to compute {\sf isa'} we use FQL's ``select/from/where'' syntax;
 an example of this syntax is shown in Figure~\ref{fql}.  Note that FQL's select/from/where syntax is syntactic sugar: the select/from/where syntax is equivalent to a data migration of the form $\Sigma \circ \Pi \circ \Delta$.  
 
Now that we have the {\sf isa'} relation ($T-$instance) on Portal A's vocabulary, we enrich Portal A's data by joining it and the {\sf isa'} relation together.  Conceptually, the enrichment process is similar to the process where ${\sf isa} \subset O \times O$ was enriched by ${\sf syn} \subset O \times N$, resulting in ${\sf isa'} \subset N \times N$; however, because Portal A's schema is not a simple relation schema, it is impractical to write the FQL code for the enrichment by hand, even using FQL's select/from/where syntax.  Hence, we developed an FQL extension to generate the required FQL code from the definition of Portal A's schema.  The result of enrichment is a new, larger instance on Portal A's schema.

\newpage
\subsection{Step 3: Query processed data}
\begin{figure}
\begin{center}
\includegraphics[width=4.8in]{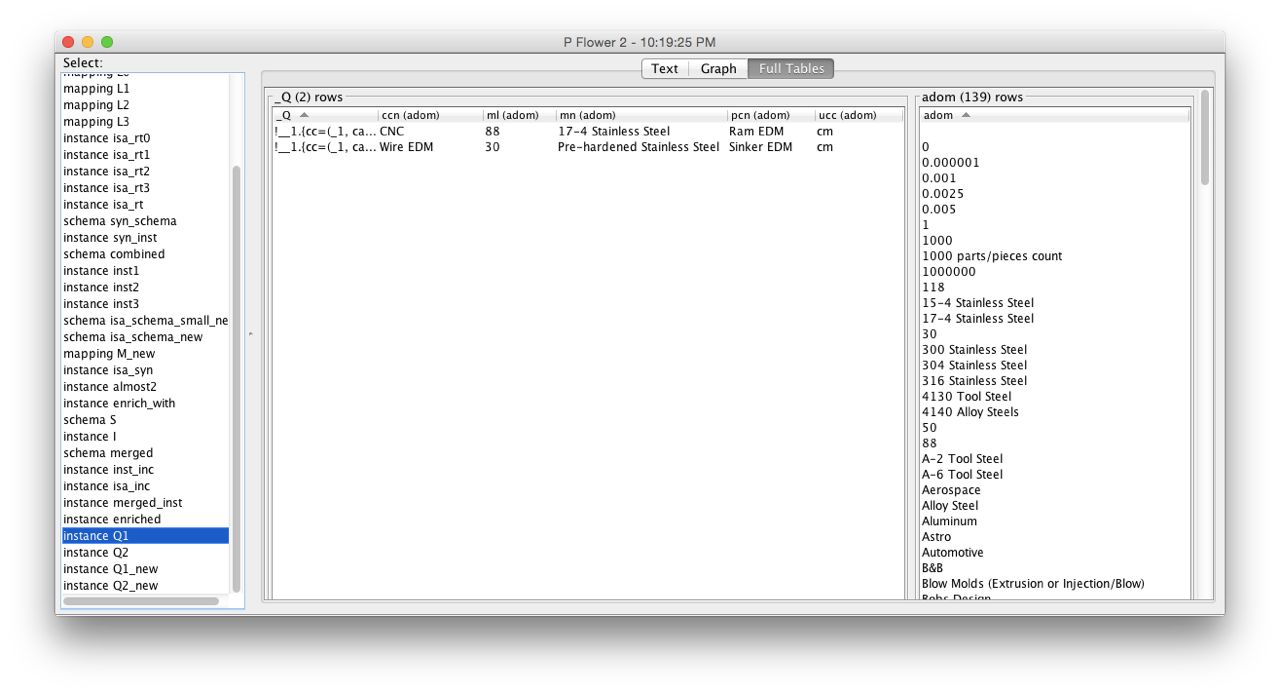}
\end{center}
\vspace{-.4in}
\caption{Query result on initial data displayed in the FQL IDE}
\label{q1old}
\end{figure}

\begin{figure}
\begin{center}
\includegraphics[width=4.8in]{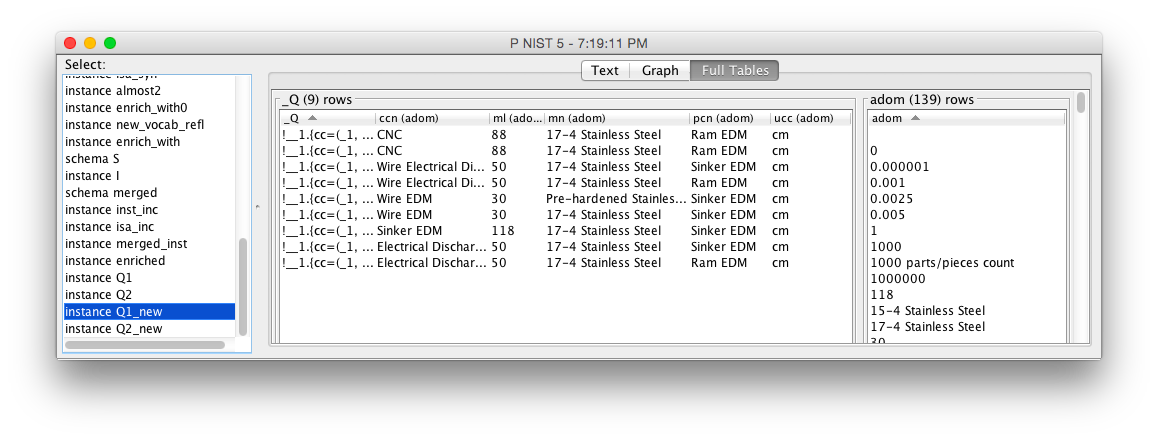}
\end{center}
\vspace{-.4in}
\caption{Query result on enriched data displayed in the FQL IDE}
\label{q1new}
\end{figure}

Having enriched Portal A's data, we can query it, using query 1 from~\cite{serm}.  The query we are using is written in FQL's select/from/where syntax and is shown in Figure~\ref{fql}.  Before enrichment, this query returns only two rows (Figure~\ref{q1old}).  After enrichment, this query returns many more rows (Figure~\ref{q1new}), because the {\sf isa'} relation contains many kinds of pre-hardened stainless steel.  (Note that ``Pre-hardened stainless steel'' does not appear in Figure~\ref{function} because that term is used by Portal A but not by the 3rd party OWL materials ontology). 


\section{Conclusion}
By implementing this example we have shown that FQL can express ``semantic enrichments'' similar to those described in~\cite{serm}.  However, this particular example is innately relational: all of the data migrations described in this paper can be implemented in SQL, albeit more verbosely than in FQL.  One promising direction for future work is to implement in FQL an enrichment scenario from~\cite{serm} whose semantics cannot be expressed in SQL, although this will require both an understanding of the relationship between OWL and FQL and a  formalization of what exactly the ``black-box OWL reasoner'' employed in ~\cite{serm} is doing.  The FQL IDE includes several information integration examples that cannot be expressed in SQL, but they are smaller than the example described in this paper and they are not ``enrichments'' in the sense of~\cite{serm}.

We also learned a valuable lesson in functorial query language design and implementation by developing this example.  Not only does FQL's select/from/where query syntax save time and effort compared to writing $\Sigma \circ \Pi \circ \Delta$ migrations, in many cases we were able to write select/from/where queries when we had no idea how to write the corresponding $\Sigma \circ \Pi \circ \Delta$ migration.  Moreover, FQL's select/from/where queries can  be executed directly in a more efficient manner than by translation to a migration of the form $\Sigma \circ \Pi \circ \Delta$.  The reason is that many techniques from relational database theory, such as join re-ordering, can be applied directly to select/from/where syntax. Hence we conclude that select/from/where syntax should be primitive in any functorial query language.  The mathematical foundations of select/from/where queries are described in~\cite{fpql}.  
  
{\bf Disclaimer.} Mention of commercial products or services in this paper does not imply
approval or endorsement by NIST, nor does it imply that such products or services are necessarily the best available for the purpose.

\bibliographystyle{plain}

\begin{figure}[h]
\begin{verbatim}
 select
  m.material_Material_Name as mn, 
  c.capability_Capability_Name as ccn,
  c.capability_Max_Length as ml,  
  uc.unitcode_Code as ucc,
  posc.productorservicecategory_Category_Name as pcn
 from
  productorservicecategory as posc, 
  material as m,  
  unitcode as uc, 
  capability as c, 
  capabilitymaterials as cmX,
  capabilitycategories as cc
 where
  c = cmX.capabilitymaterials_Capability_id and
  uc = c.capability_Max_Length_Unit and 
  uc.unitcode_Code="cm" and
  m = cmX.capabilitymaterials_Material_id and
  c = cc.capabilitycategories_Capability_id and
  posc = cc.capabilitycategories_ProductOrServiceCategory_id and
  (m.material_Material_Name="Pre-hardened Stainless Steel" or 
   m.material_Material_Name="17-4 Stainless Steel") and
  (posc.productorservicecategory_Category_Name="Sinker EDM" or 
   posc.productorservicecategory_Category_Name="Ram EDM")
\end{verbatim}
\caption{FQL syntax for Query 1~\cite{serm}, translated from SQL}
\label{fql}
\end{figure}


\end{document}